\documentclass{aa}  

\usepackage{mathabx}
\usepackage{graphicx}
\usepackage{txfonts}

\begin{document}


   \title{Formation of terrestrial planets in eccentric and inclined giant planet systems}


   \author{Sotiris Sotiriadis \inst{1}
          \and
          Anne-Sophie Libert  \inst{1} 
          \and
          Sean N. Raymond     \inst{2}
          }

   \institute{
         naXys, Department of Mathematics, University of Namur, 8 Rempart de la Vierge, 5000 Namur, Belgium\\
         \email{sotiris.sotiriadis@unamur.be}
         \and
         Laboratoire d’Astrophysique de Bordeaux, Université de Bordeaux, CNRS, B18N, Allée Geoffroy Saint-Hilaire, 33615 Pessac, France
             }

   \date{Received ....................; accepted ...............}

  \abstract                        
   {}
   {Evidence of mutually inclined planetary orbits has been reported for giant planets these last years. Here we aim to study the impact of eccentric and inclined massive giant planets on the terrestrial planet formation process, and investigate whether it can possibly lead to the existence of inclined terrestrial planets.} 
   {We have performed 126 simulations of the late-stage planetary accretion in eccentric and inclined giant planet systems. The physical and orbital parameters of the giant planet systems result from n-body simulations of three giant planets in the late stage of the gas disc, under the combined action of Type II migration and planet-planet scattering. Fourteen two- and three-planet configurations have been selected, with diversified masses, semi-major axes (resonant configurations or not), eccentricities and inclinations (including coplanar systems) at the dispersal of the gas disc. We have then followed the gravitational interactions of these systems with an inner disc of planetesimals and embryos ($9$ runs per system), studying in detail the final configurations of the formed terrestrial planets.}
   {Besides the well known secular and resonant interactions between the giant planets and the outer part of the disc, giant planets on inclined orbits also strongly excite the planetesimals and embryos in the inner part of the disc, through the combined action of nodal resonance and Lidov-Kozai mechanism. This has deep consequences on the formation of terrestrial planets. While coplanar giant systems harbour several terrestrial planets, generally as massive as the Earth and mainly on low eccentric and low inclined orbits, terrestrial planets formed in systems with mutually inclined giant planets are usually fewer, less massive ($<0.5 M_{\Earth}$) and with larger eccentricities and inclinations. This work shows that terrestrial planets can form on stable inclined orbits through the classical accretion theory, even in coplanar giant planet systems emerging from the disc phase.} 
   {} 

   \maketitle
   

\section{Introduction}
	Observations have revealed various orbital properties of extrasolar systems. In particular, according to the radial velocity surveys, giant planets appear with diversified eccentricity distribution, while the existence of strongly misaligned "hot-Jupiters" gives some clues about non-coplanar systems \citep{2010A&A...524A..25T,Albrecht1}. Several $n$-body studies have shown that the formation of eccentric and inclined giant planet systems can result from different mechanisms. Planetary migration can lead to eccentricity and inclination resonant excitation \citep{Lee_Peale,Thommes2003b, Libert2009} and planet-planet scattering, after the dispersal of the disc, can also reproduce the eccentricity distributions of the detected gas giants and produce orbits with high inclination, if the systems are initially in unstable configurations \citep{Marzari1996,Rasio1996,Adams2003,Raymond2008,Chatterjee2008,Juric,Ford2008,2010ApJ...711..772R,Petrovich2014}. However, these initial conditions do not take into account the imprint of the disc era \citep{Lega2013}.

	More realistic scenarios have investigated the gravitational interactions among the planets during the migration phase and highlighted that dynamical instabilities in multi-planet systems can occur throughout the gas phase \citep{Moorhead2005,2008ApJ...688.1361M,Marzari2010,Matsumura2010,Libert2011a,Moeckel2012}. Recently, \cite{Sotiriadis2017} have realized extensive n-body simulations of three giant planets in the late stage of the disc, considering disc-induced Type-II migration and the damping formulae for eccentricity and inclination fitted from the hydrodynamical simulations of \cite{Bitsch2013}. While their simulations reproduce the semi-major axis and eccentricity distributions of the detected giant planets, they have also shown that highly mutually inclined systems can be formed despite the strong eccentricity and inclination damping. A detailed analysis of the long-term dynamical evolutions of these systems is also provided, with a particular emphasis on the different inclination-growth mechanisms.  

	After the protoplanetary disc phase (a few Myr) during which giant planets have accreted their gaseous envelopes, they affect gravitationally the remaining swarm of solid planetesimals and the tens to hundreds planetary embryos across the system \citep{Kokubo1998,Thommes2003a}. The so-called post-oligarchic growth phase is the last phase of the terrestrial planet formation process and occurs on a timescale of $10^7$-$10^8$ years \citep{Chambers2001_t,Raymond2004,Raymond2005, OBrien2006,2010Icar..207..517M}. The already formed giant planets have an essential and important impact on this process and eventually on the final long-term architecture of the planetary system (see e.g. \citealt{Morbi_review} and \citealt{PPVI_Raymond} for a review on terrestrial planet formation). 

	The influence of giant planets' eccentricity on terrestrial accretion has been studied by several authors. Eccentricity excitation of planetary embryos due to gravitational perturbations by outer giants seems to be the most crucial part of the late-accretion phase \citep{Chambers2002, LevisonAgnor,Raymond2006}. As might be expected, there is a correlation between the scale of these perturbations and the orbital characteristics of the giants, especially their mass and eccentricity. \citet{LevisonAgnor} have noted that giants on eccentric orbits remove from the system a large fraction of embryos and consequently less terrestrial planets are formed, usually on orbits with larger eccentricities. In addition, they have shown that planets tend to form closer to the star if the giants, exterior to the disc, are more eccentric. 
	
	Moreover, planet-planet scattering, following dynamical instabilities in systems with multiple gas giants, can have catastrophic effects on terrestrial formation. During the instability period, planetesimals and embryos could either be driven to the central star or be scattered in very eccentric orbits and eventually be ejected out of the system, making terrestrial accretion inefficient \citep{Veras2005,Veras2006}. \citet{Raymond2011,Raymond2012} have investigated the formation of terrestrial planets under the influence of both stable and unstable planetary systems with three gas giants. They have noted that an anti-correlation exists between the eccentricity of the innermost giant and the total mass of the terrestrial planets. They have also pointed out that it is common for the formed terrestrial planets to survive in eccentric and inclined orbits, especially for single-terrestrial planets. \citet{Matsumura2013} have shown that only the terrestrial planets very close to the parent star could survive in three-giant systems with high eccentricities. \citet{Carrera2016} have highlighted that it is extremely difficult for habitable terrestrial planets to survive in systems of three Jupiter-like planets that suffer instabilities. They have also shown that the probability to survive and remain habitable is higher for giants evolving on orbits with larger semi-major axes and lower eccentricities.

 \begin{figure}[t!]	
       \centering
       \includegraphics[width=1.0\hsize]{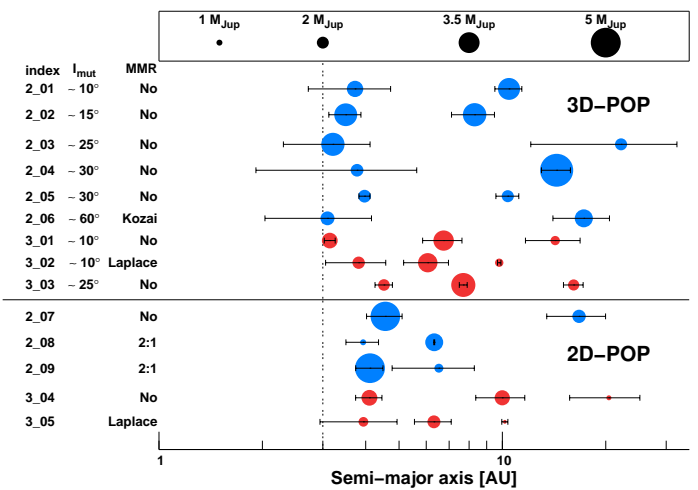} 
       \caption{Initial giant planetary configurations considered in this work. The systems of 3D-POP includes two planets whose mutual inclination is larger than $10^{\circ}$ (see the second column for the different values of mutual inclination considered), while all systems of 2D-POP are coplanar. Two-planet systems are shown in blue, three-planet ones in red, the size of the points varying with the cubic root of the planetary mass (see the scale on top). The horizontal axis shows the semi-major axis of the planets in logarithmic scale. The error-bars represent the apastron and periastron of the planets. On the left panel, the first column corresponds to the index of each system, and the third one indicates if the system is in a resonant configuration. The vertical dashed line shows the outer edge of the disc of planetesimals and embryos.}
       \label{fig:fig1}
 \end{figure}

	Concerning giant planets on inclined orbits, \citet{LevisonAgnor} has considered the impact of three slightly inclined planets on the terrestrial planet formation and highlighted that the excitation of the embryos in a region of the disc can be transferred to another region (secular conduction). Also, simulations by \citet{Jin2011} have investigated the late stage of terrestrial accretion  
	in the system OGLE-2006-BLG-109L, considering several inclination values of the outer giant. They have shown that terrestrial planets can be formed, even inside the habitable zone, and that the effectiveness of embryo accretion drops as the relative inclination of the two giant planets increases.

	In the present work, we aim at studying the terrestrial planet formation in systems consisting of two or three massive giant planets, exterior to the initial disc of solids, that are on eccentric and highly mutual inclined orbits. Our goal is to examine whether (or not) planets could emerge from such configurations through the classical accretion theory and especially if terrestrial planets could be formed on inclined orbits by this mechanism. The physical and orbital parameters of the giant planet systems result from n-body simulations of three giant planets in the late stage of the gas disc, under the combined action of Type II migration and planet-planet scattering, following \citet{Sotiriadis2017}.

	We stress that our survey consists in ``case studies", in the spirit of \citet{LevisonAgnor}, \citet{Raymond2011,Raymond2012} and \citet{Jin2011}, rather than a systematic exploration of the parameter space, similar to \citet{Chambers2002}, \citet{Raymond2004,Raymond2009}, \citet{OBrien2006}, \citet{2010Icar..207..517M}, \citet{FISCHER201428} and \citet{2015Icar..252..161K} focusing mainly on the Solar System. The advantage of this approach is that the case studies each have a full, realistic dynamical history with regard to the giant planet-disc late interactions, and do not adopt somehow arbitrary initial conditions for the giants. A limitation of this approach is that, compared with a systematic study, many parameters change at once, and so it is not trivial to determine exactly the influence of each parameter. In that sense, our approach here is complementary to the systematic approach because, while systematic studies show how different giant planet orbits affect terrestrial planet formation, here we focus only on parts of the giant planet parameter space naturally populated by early dynamical evolution.
	
	In Section~\ref{section2}, we describe the set-up of our numerical experiments and the parameters of the giant planet systems considered in our work. Typical outcomes of our simulations are presented in Section~\ref{section3}, and the impact of inclined giant planets on the disc of planetesimals and embryos is studied in detail in Section~\ref{section4}. In Section~\ref{section5}, we describe the physical and orbital parameters of the terrestrial planets formed in our simulations. Finally, our conclusions are given in Section~\ref{section6}.

\section{Methods}\label{section2}

	In the present study, we investigate the formation of terrestrial planets in 14 planetary systems, consisting of two or three giant planets around a solar-mass star. The systems are followed during the post-oligarchic growth phase \citep{Kokubo1998,Thommes2003a}, also known as late-stage accretion, where terrestrial planets emerge from accretion of embryos and planetesimals. Indeed, planetary embryos on eccentric orbits no longer have independent feeding zones but, due to orbit crossings, collide with other embryos and planetesimals. Eccentricity growth of the embryos and eventually the efficiency of terrestrial planet accretion strongly depend on the orbital configuration of the giant planets that exist in the system. The parameters of the giant planet systems are described in Section \ref{subgiant} and the set-up of our numerical experiments in Section \ref{subterr}.

	Let us note that the set-up of our simulations is based on \citet{Raymond2011}. However, the major difference is that, in our study, we do not consider arbitrary initial conditions for the giant planets, but their orbits carry the imprint of the protoplanatery disc phase. Indeed, we have followed the evolution of the giant planets in the late stage of the gas disc, taking into account planet-planet interactions and disc-planet interactions. We consider here the configurations of the giant planetary systems as they emerged from the disc phase.

\subsection{Architecture of the giant planet systems}\label{subgiant}	
	
	The physical and orbital parameters of the giant planet systems considered in our study are shown in Figure~\ref{fig:fig1}. Instead of arbitrary initial conditions for the giant planets, we have followed \cite{Sotiriadis2017}, where both the combined action of the gas disc (Type II migration and eccentricity/inclination damping) and the planet-planet interactions are taken into account, to set up the architecture of the giant planet systems. In particular, we have run 300 n-body simulations of three giants on quasi-circular and quasi-coplanar orbits ($e \in [0.001,0.01]$ and $i \in [0.01^{\circ},0.1^{\circ}]$) in the late stage of the gas disc. The initial semi-major axis of the inner planet is fixed to $5$ AU, while the middle and outer ones follow uniform distributions in the intervals $a_{2}\in[7.25,10.75]$ AU and $a_{3}\in[13,25]$ AU, respectively. These initial distances to the star are such that the formation of terrestrial planets can occur around $1$ AU. We choose randomly initial planetary masses from a log-uniform distribution in the interval $[0.65,5] M_{\rm Jup}$. Through the evolution of the system, we decrease the disc mass exponentially, with a dispersal time of $1$ Myr, and let the simulations run until $1.4$ Myr. 
	
	Among these hundreds of giant systems, we select $9$ representative two- and three-planet configurations with at least one pair of planets having a high mutual inclination ($I_{mut}\,\gtrsim\,10^{\circ}$)\footnote{$ \cos I_{mut} = \cos I_{1} \cos I_{2} + \sin I_{1} \sin I_{2} \cos(\Omega_{2} - \Omega_{1}).$}, referred to as 3D-POP in the following, and $5$ representative two- and three-planet coplanar configurations, called 2D-POP. The architecture of the $14$ systems are depicted in Figure~\ref{fig:fig1}. The horizontal axis corresponds to the semi-major axes of the planets and the errorbars represent their apastron and periastron, reflecting the eccentricity of the orbit. In most cases, the eccentricities are moderate to high, and half of the systems host a planet with an eccentricity larger than $0.25$. The vertical dashed line, at $3$~AU, sets the limit for the outer edge of the disc. The size of each circle is proportional to the cubic root of the planetary mass and the frame on top of the figure shows four different masses for scale. Systems of two planets and three planets are colored in blue and red circles, respectively. Several resonant configurations are considered here, namely the 2:1 mean motion resonance, the Laplace resonance (4:2:1 resonance, similar to the one in the Galilean moons) and the secular Lidov-Kozai mechanism \citep{Lidov,Kozai}. The index of each system (labelled from \texttt{2\_01} to \texttt{2\_09} and from \texttt{3\_01} to \texttt{3\_05}), its mutual inclination if inclined, and its resonant configuration if any, are given in the three columns at the left of the figure. In summary, a variety of physical and orbital parameters of the planets (different mass ratios, orbital separations, eccentricities, mutual inclinations, and resonances) are considered here, in order to see their impact on the terrestrial planets that will be formed in our simulations. 

 \begin{figure}
       \centering
       \includegraphics[width=\hsize]{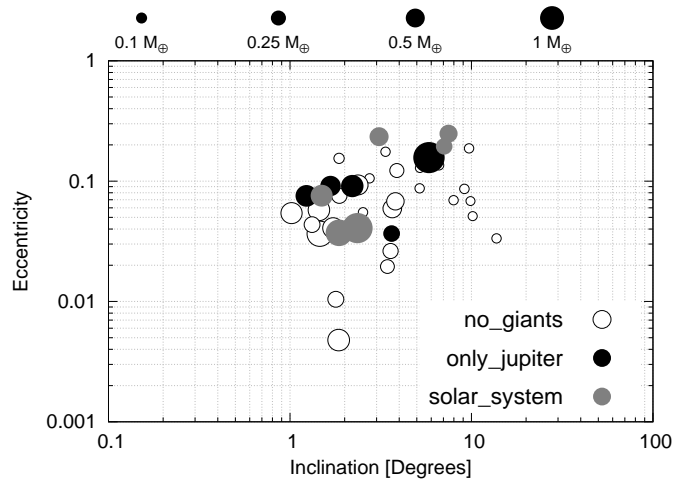} 
       \caption{Eccentricity versus inclination of the terrestrial planets for the three reference cases. The white, black and grey dots correspond to the \texttt{no\_giants}, \texttt{only\_Jupiter} and \texttt{Solar\_System} set-ups, respectively. The size of each circle is proportional to the cubic root of the planetary mass.}
       \label{new2}
 \end{figure}
	
 \begin{figure}
       \centering
       \includegraphics[width=0.49\hsize]{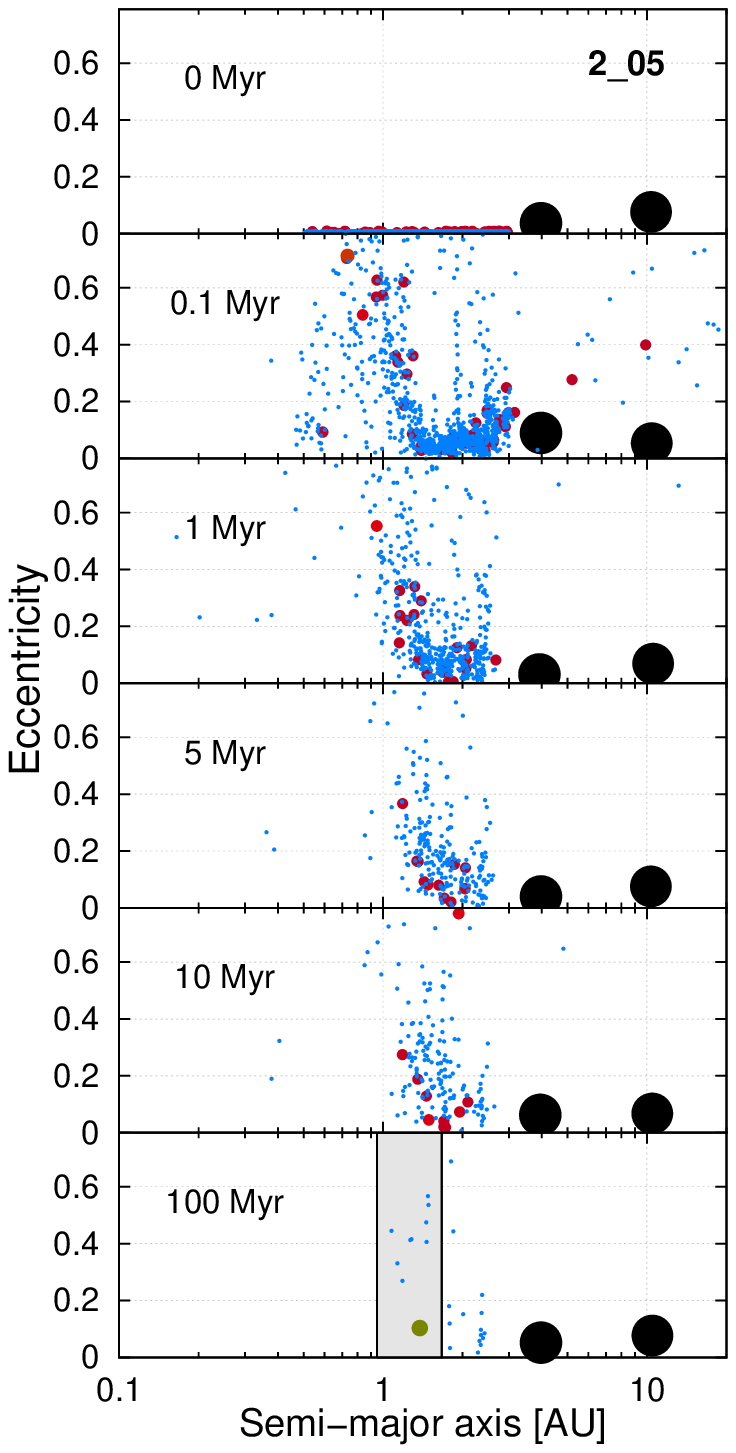}
       \includegraphics[width=0.49\hsize]{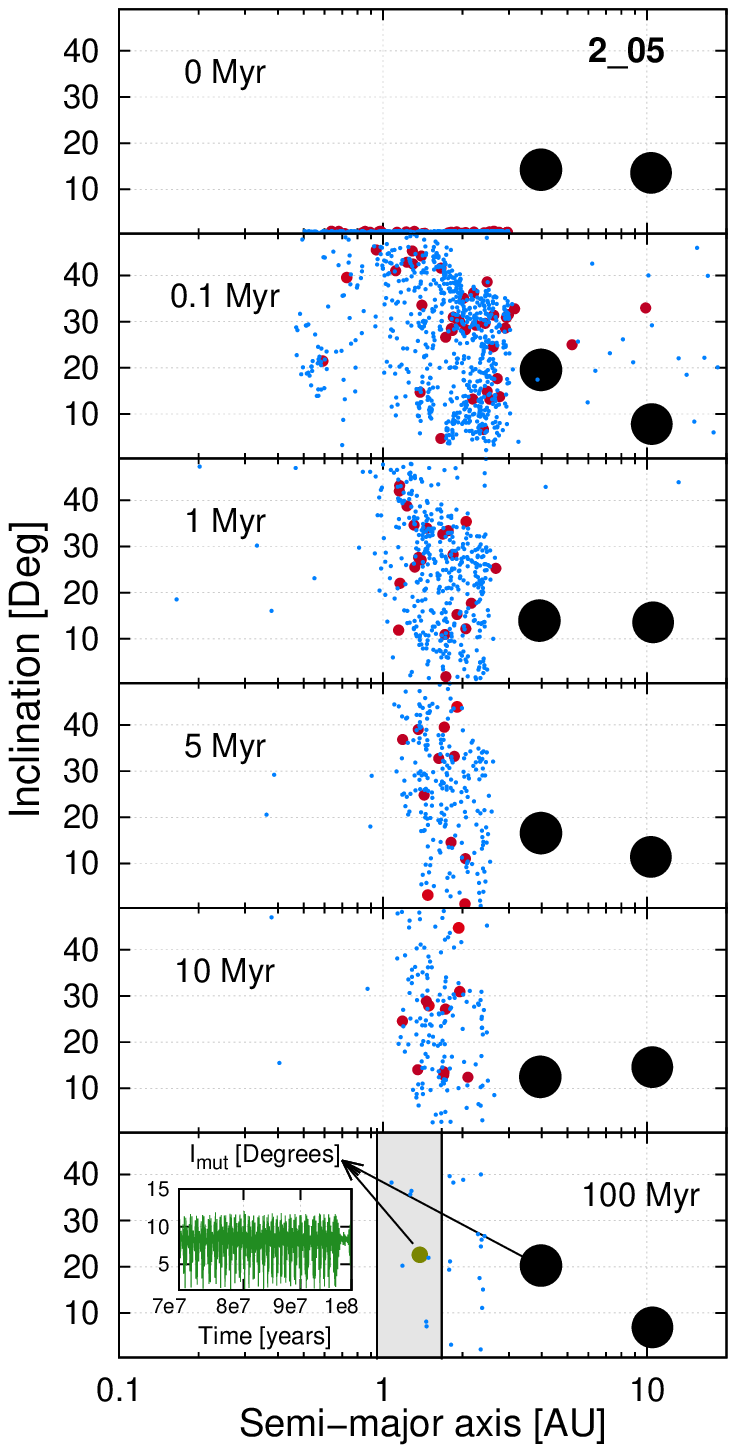}
       \includegraphics[width=0.95\hsize,keepaspectratio]{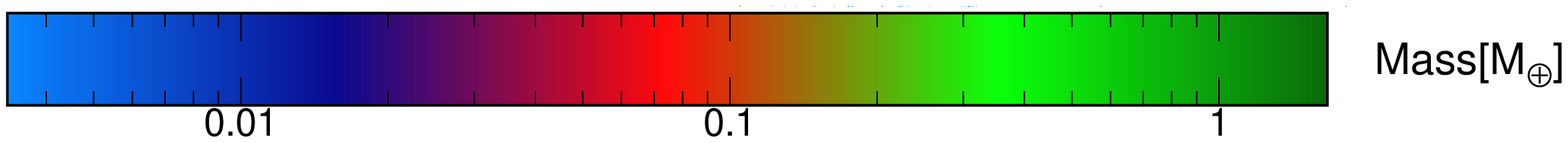}
       \caption{Snapshots in time from a non-coplanar system (3D-POP, \texttt{2\_05}) with $I_{mut}\!\approx\!30^{\circ}$ initially. Evolution of the eccentricities is shown in left, and the inclinations in right. The size of each circle is proportional to the cubic root of its mass (
       see also the colorbar). The giant planets are displayed with black circles. The light-grey shaded region corresponds to the habitable zone. The evolution of the mutual inclination between the inner giant and the largest terrestrial body in the last $30$ Myr is shown in the inset plot.}
       \label{fig:case1}
 \end{figure}

 \begin{figure}
       \centering
	\includegraphics[width=0.527\hsize]{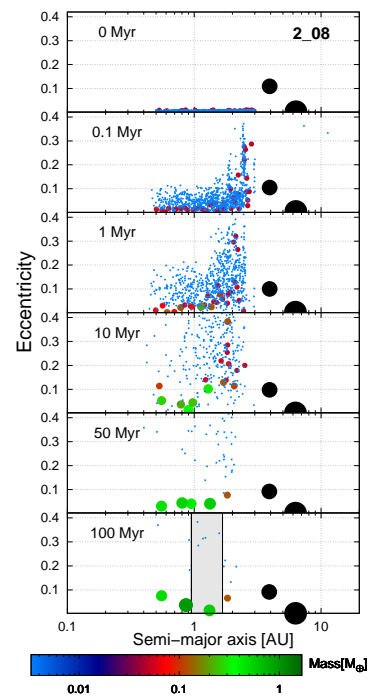} 
	\caption{Snapshots in time from a coplanar system (2D-POP, \texttt{2\_08}) in a 2:1 mean-motion resonant configuration. At $100$ Myr, four terrestrial planets are formed in a configuration similar to the Solar System. Formatted as in Fig. \ref{fig:case1}.}
       \label{fig:case2}
 \end{figure}

 \begin{figure}
       \centering
       \includegraphics[width=0.49\hsize]{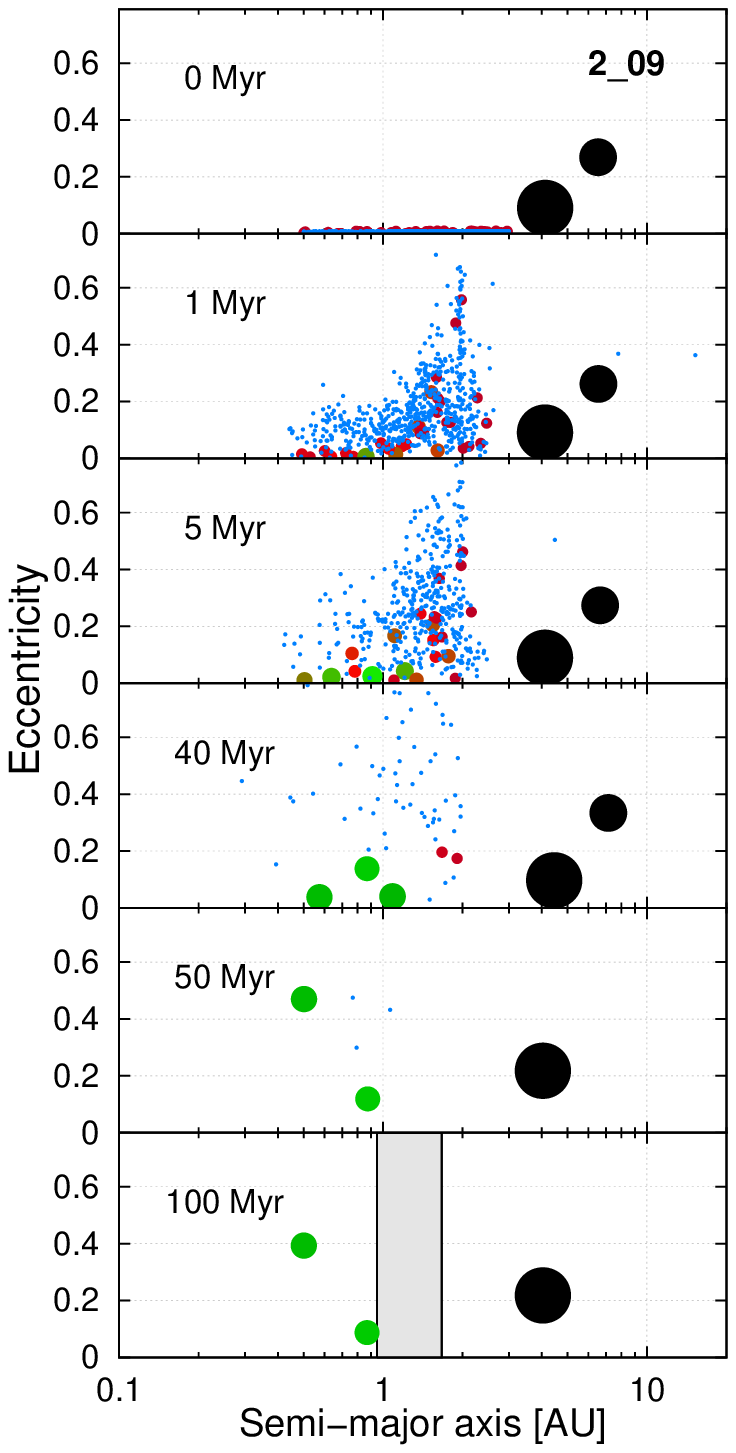}
       \includegraphics[width=0.49\hsize]{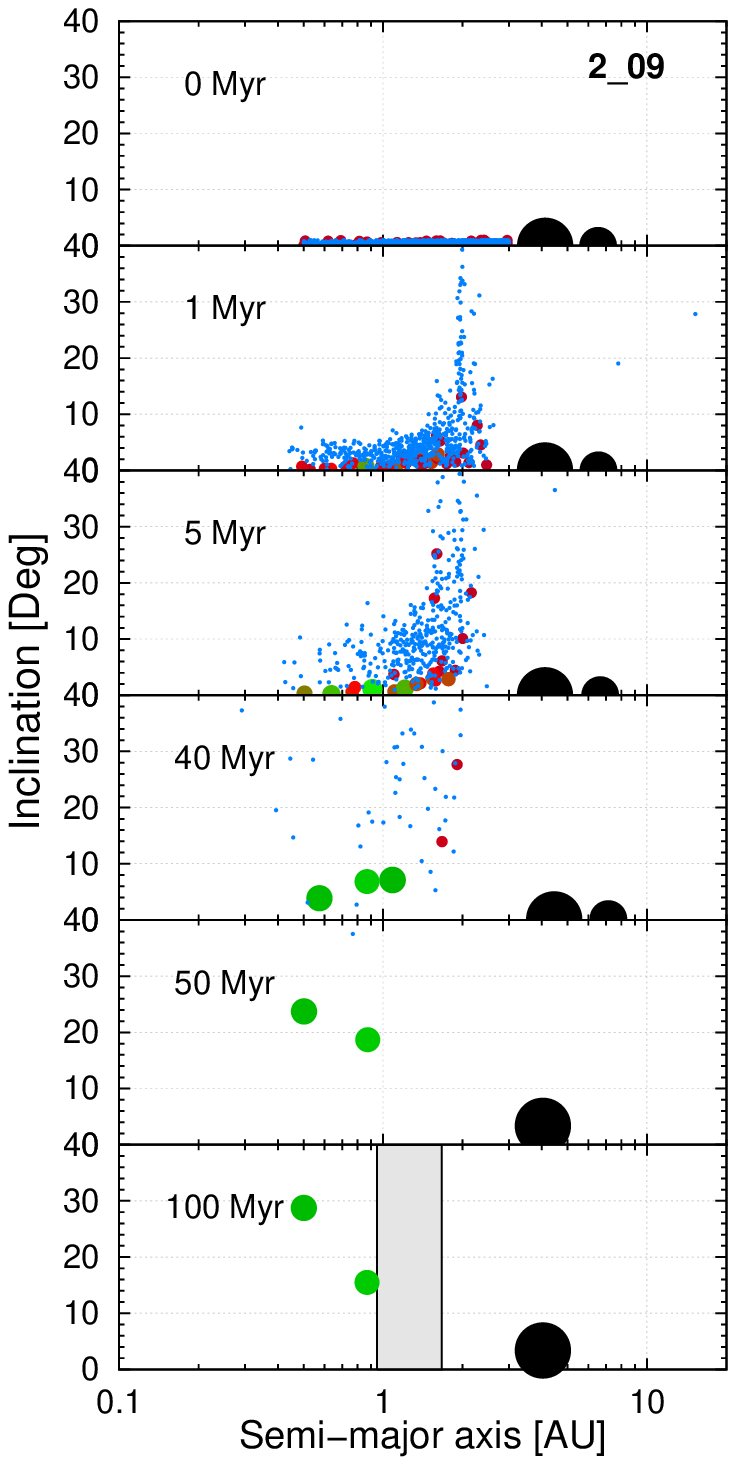}
       \includegraphics[width=0.95\hsize,keepaspectratio]{figures/colorbar.eps}
       
       \caption{Snapshots in time from a coplanar system (2D-POP, \texttt{2\_09}) with initially moderate eccentricities. At $40$ Myr, three terrestrial planets are formed and due to gravitational interactions, the outer giant planet is ejected from the system at $\sim\!47$ Myr. Only two Earth-mass planets survive the destabilisation phase, and remain on eccentric and inclined orbits until the end of the simulation. Formatted as in Fig. \ref{fig:case1}.}
       \label{fig:case3}
 \end{figure}

\subsection{Description of the simulations}\label{subterr}	

	As already mentioned, we follow a similar set-up for the terrestrial disc as in \citet{Raymond2011}, except that we do not take into account an outer planetesimal disc. We consider a disc of solids lying between $0.5$ and $3$ AU and consisting of a swarm of planetesimals and planetary embryos. The disc's surface density follows a flat radial profile $\Sigma_{\rm solids}(r)~\propto~r^{-1} $ and the internal density for all the planetesimals and embryos is $\rho = 3$ gr/cm$^{3}$. The disc initially consists of $50$ embryos, and $1000$ planetesimals that do not interact gravitationally with each other but only with the embryos and the planets, in order to decrease the computational cost \citep{Raymondetal2006}. The embryos are slightly less massive than Mercury, $m_{\rm emb}=0.05 M_{\Earth}$, and the planetesimals are as massive as Pluto, $m_{\rm pl} = 0.0025 M_{\Earth}$. The total mass of the disc is thus $M_{\rm disc} = 5 M_{\Earth}$ and the total mass ratio for embryos and planetesimals is $M_{\rm emb}/M_{\rm pl}=1$ \citep{2006AJ....131.1837K}. The embryos are spaced, in terms of mutual Hill radii, by $K \approx 7\!\!-\!\!8~R_{H,m}$, where 
\begin{equation}
	R_{H,m} = \left(\frac{m_{1}+m_{2}}{3M_{\rm star}}\right)^{1/3}~\left(\frac{a_1+a_2}{2}\right).
\end{equation}
	The initial eccentricities and inclinations of the solids are chosen randomly from a uniform distribution in the ranges $[0.001,0.01]$ and $[0.01^{\circ},1^{\circ}]$, respectively. Doing so, the initial terrestrial disc plane coincides, in our simulations, with the one of the gas disc in the late protoplanetary disc phase (see Section~\ref{subgiant}). As already mentioned, our work focuses on the late-stage accretion phase, so we assume that there is no gas/dust disc left in the systems and the gas giants are fully formed. Of course, the late-gas phase and the late-accretion phase are not independent of each other, and the interactions between the two phases are not taken into account in this work for computational reasons (see the discussion on the limitations of our model in Section~\ref{section6}). 

	To perform the $n$-body simulations, we use the symplectic integrator SyMBA \citep{symba}, which handles close encounters between the bodies\footnote{The term 'bodies' here refers to planets, embryos and planetesimals.} by using a multiple time step technique. Moreover, due to the highly eccentric and inclined configurations of the giants, embryos and planetesimals are excited in very eccentric orbits and this means that high resolution is also required for close encounters between the bodies and the star. For this reason, we adopt a symplectic algorithm that has the desirable property of being able to integrate close perihelion passages with the parent star \citep{Levison1}. Nine runs are performed for each of the $14$ configurations of giant planet systems, each one with a different randomly generated disc. The systems are integrated up to $100$ Myr and our time-step is fixed to $dt=0.01$ yrs. We treat the possible merging between two bodies as a totally inelastic collision, when their distance becomes less than the sum of their radius. The boundary value for accretion onto the star is 0.01 AU and the one for ejection from the system, 1000 AU. The computational effort required for our investigation is $\sim~5\times 10^4$ computational hours.

	\subsection{Reference cases} \label{primary}	
	For future comparison, three additional sets of simulations (each one consisting of two runs) have been performed. In the \texttt{no\_giants} set-up, the disc of planetesimals and embryos is evolved without any giant. In the \texttt{only\_Jupiter} set-up, the disc of planetesimals and embryos is affected by a giant planet with the mass and orbital elements of Jupiter. We investigate the terrestrial accretion when considering the four gas/ice giants in their current orbit in the \texttt{Solar\_System} set-up.
	
	The orbital parameters of the terrestrial planets with mass~$>0.05M_{\Earth}$ formed in these three sets of simulations are displayed in Fig.~\ref{new2}, after 100 Myr. The size of each circle is proportional to the cubic root of the planetary mass. Altough more bodies with slightly diversified parameters are formed in the \texttt{no\_giants} simulations, the parameters of the terrestrial planets formed in the three sets show similar features: all the bodies have eccentricity smaller than $0.3$ and inclination smaller than $10^\circ$ at the end of the simulation. Moreover, we note the formation, in the three sets, of several terrestrial planets with a mass comparable to the Earth or slightly more massive. 
		

 \begin{figure}
       \centering
       \includegraphics[width=0.49\hsize]{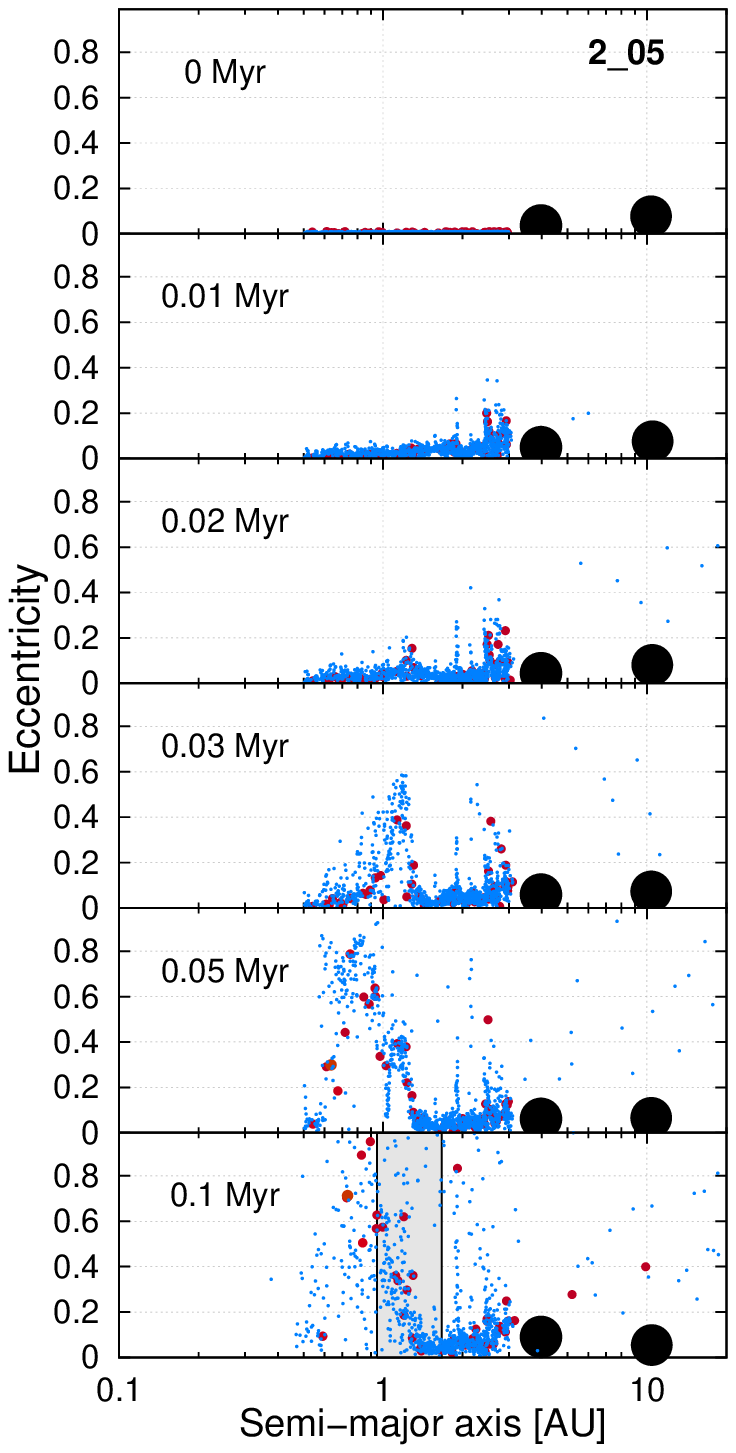}
       \includegraphics[width=0.49\hsize]{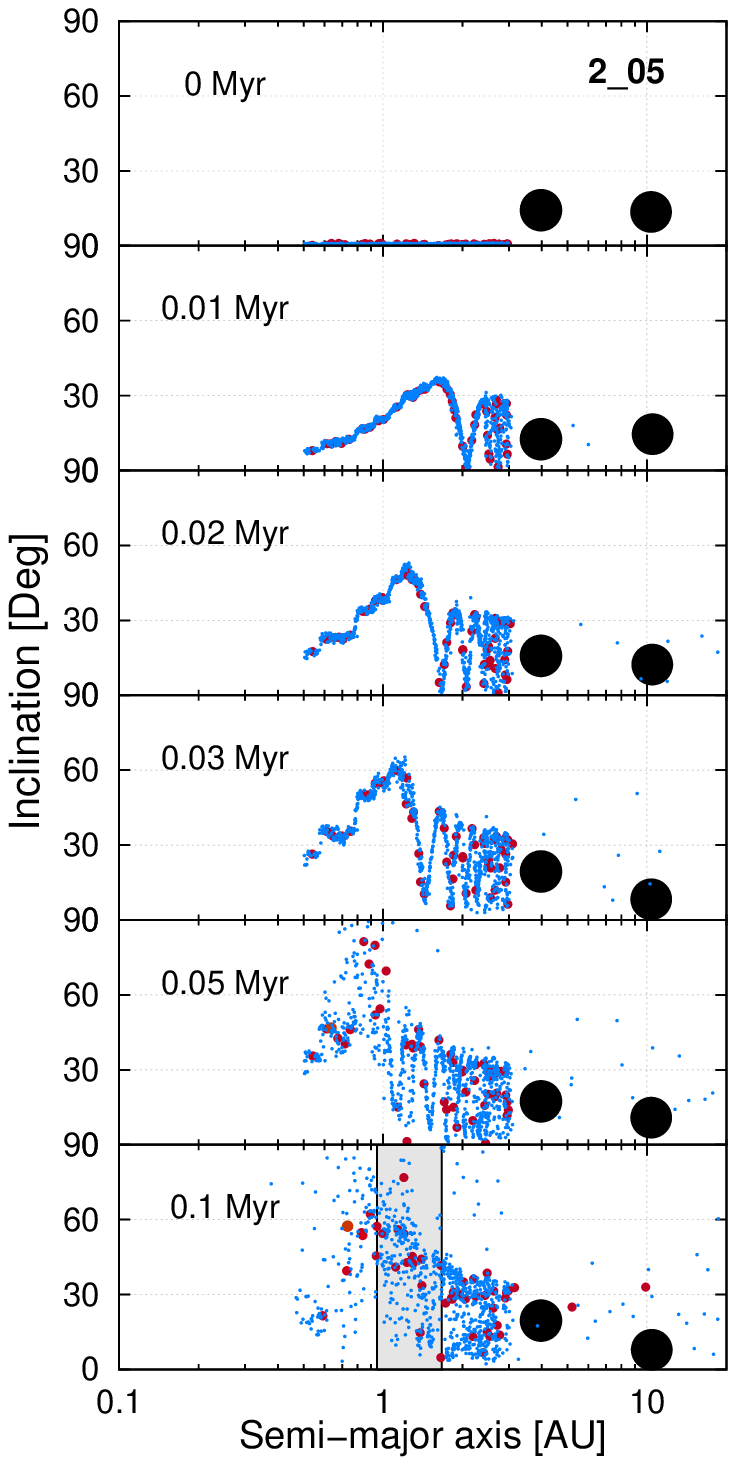}
       \includegraphics[width=0.95\hsize,keepaspectratio] {figures/colorbar.eps}
       \caption{Snapshots in time of the first $0.1$ Myr of the non-coplanar system (3D-POP, \texttt{2\_05}) presented in Fig.~\ref{fig:case1}.  The eccentricity and inclination waves are discussed in the text.}      \label{fig:Kozai1}
 \end{figure}

 \begin{figure}
       \centering
       \includegraphics[width=0.75\hsize]{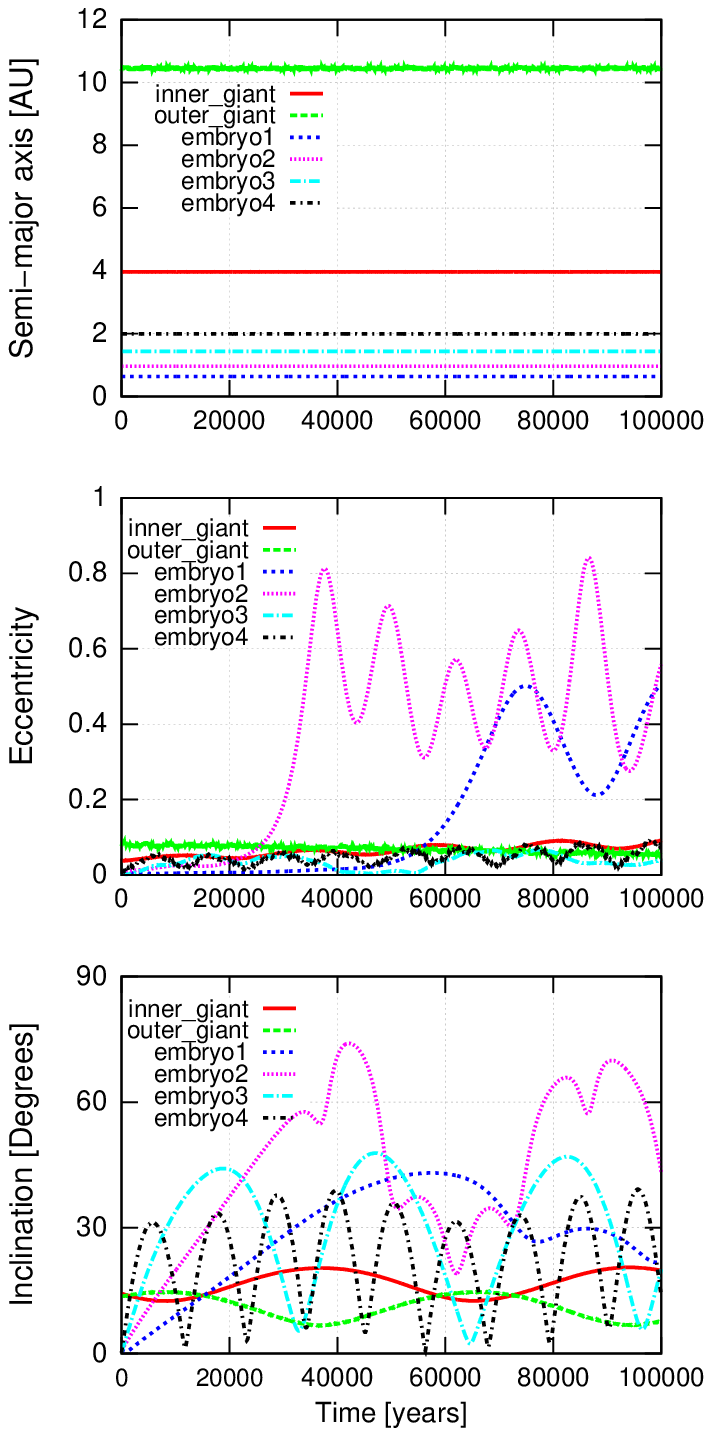} 
       \caption{Evolution of four embryos in the \texttt{2\_05} giant planet architecture, during the first $0.1$ Myr. Large variations in eccentricity and inclination are observed, due to nodal resonance and Lidov-Kozai mechanism. See text for more details.}
       \label{fig:Kozai2}
 \end{figure}

\section{Typical evolutions}\label{section3}
	In this section, we describe three representative outcomes of our simulations of late-stage planetary accretion in the giant planetary configurations shown in Fig. \ref{fig:fig1}. They illustrate the dynamical excitation of the planetesimals and embryos by the gas giants, the subsequent possible rearrangement of the giants, as well as the properties of the terrestrial planets that are formed.      
		
	In Fig.~\ref{fig:case1} we show the interaction of a non-coplanar system (3D-POP, \texttt{2\_05}) with the inner disc of planetesimals and embryos. We present six snapshots in time of the evolution of the eccentricities (left panels) and inclinations (right panels) of each body, at $t=0, 0.1, 1, 5, 10$ and $100$ Myr. The size of a circle is proportional to the cubic root of the mass of the terrestrial body, and a colorscale is added for clarity. The two giant planets (black circles), whose masses are $m_{\rm in}=2.15 M_{\rm Jup}$ and $m_{\rm out}=2.02 M_{\rm Jup}$, are initially on highly inclined orbits with $I_{\rm mut} \approx 30^{\circ}$, and the inner giant is relatively close to the disc ($a_{\rm in} \approx 4$ AU). The two giant planets will keep their inclined configuration throughout the whole evolution of the system. As can be observed from the snapshot at $0.1$ Myr, planetesimals and embryos are very early strongly excited, both in eccentricity and inclination. An in-depth study on the dynamical excitation of the disc by inclined giant planets is realized in Section~\ref{section4}. At $100$ Myr, almost all planetesimals either have been accreted by the massive bodies (embryos, giants, star) or have been ejected from the system due to strong dynamical interactions with the giant planets. A terrestrial planet with $0.2 M_{\Earth}$ has been formed and is located inside the habitable zone (light-grey shaded area, \cite{1993Icar..101..108K,2013ApJ...765..131K}) on a stable orbit, slightly inclined with the orbital plane of the inner giant planet (see the inset plot in the last snapshot), but highly inclined with respect to the plane of the outer giant ($\sim35^\circ$). 

	For comparison, a simulation of the late-stage planetary accretion for a coplanar giant planet system (2D-POP, \texttt{2\_08}) is displayed in Fig.~\ref{fig:case2}. The giants ($m_{\rm in}\!=\!1.01 M_{\rm Jup}$ and $m_{\rm out}\!=\!3.00 M_{\rm Jup}$) are initially in a 2:1 mean-motion resonance ($a_{\rm in}\!=\!3.93$ AU, $a_{\rm out}\!=\!6.33$ AU) and remain in the resonance until the end of the simulation. This example is in line with the previous works on the late-stage formation with low-eccentric giant planets (see for instance \cite{Raymondetal2006}). Vertical spikes associated to different mean-motion resonances with the inner giant planet are clearly visible after $0.1$ Myr. While the increase of the eccentricities in the outer disc is due to secular or resonant perturbations with the giant planets, the eccentricities in the inner disc are driven by interactions between the embryos. Compared with the non-coplanar system in Fig.~\ref{fig:case1}, the terrestrial accretion is more efficient here and a Solar System analog emerges, consisting of four terrestrial planets on stable, low-eccentric and low-inclined orbits, of which one is well inside the habitable zone. 
	
	In the third evolution, we point out that terrestrial planets on inclined orbits can also form by accretion in coplanar systems, as shown by Fig.~\ref{fig:case3} (2D-POP, \texttt{2\_09}). The system consists in two giant planets with masses $m_{\rm in}= 4.95 M_{\rm Jup}$ and $m_{\rm out}= 1.52 M_{\rm Jup}$, in a 2:1 mean-motion resonance ($a_{\rm in}\!=\!4.12$ AU, $a_{\rm out}\!=\!6.53$ AU). The planets have initially moderate eccentricities: $e_{\rm in}\!=\!0.09$ and $e_{\rm out}\!=\!0.27$. Again the terrestrial accretion is very efficient in the inner disc, leading to the formation of three Earth-like planets at $40$ Myr. However, the system is rapidly destabilised due to the gravitational interactions between the bodies, leading to the ejection of the outer giant planet at $\sim\!\!47$ Myr. The scattering event produces an increase of the eccentricities and inclinations of the two residual terrestrial planets, which remain on stable eccentric and inclined orbits until the end of the simulation.

	The examples discussed here highlight that the formation of terrestrial planets on stable inclined orbits is possible through the classical accretion theory, both in coplanar and non-coplanar giant planet systems. However, we have seen that the accretion is more efficient in coplanar systems, since inclined giant planets affect more heavily the planetesimals and embryos. The dynamical mechanisms producing the excitation of the disc will be deeply analysed in the next section.      


\section{Interactions of inclined giant planets with the disc of planetesimals and embryos}\label{section4}

In this section we perform a detailed study of the strong dynamical excitation of the disc by inclined giant planets, as observed in Fig.~\ref{fig:case1} at $0.1$ Myr (3D-POP, \texttt{2\_05}). To identify the dynamical mechanisms acting at the beginning of the simulation, additional snapshots in time are provided by Fig.~\ref{fig:Kozai1}, for $0.01$, $0.02$, $0.03$, $0.05$ Myr. Besides the secular and resonant interactions between the outer disc and the giants, giving rise to the well-known vertical spikes, interesting waves in inclination in the inner edge of the disc can be observed very early in the evolution, while the eccentricities of the planetesimals and embryos remain very low. 

\begin{figure}
       \centering
       \includegraphics[width=0.48\hsize]{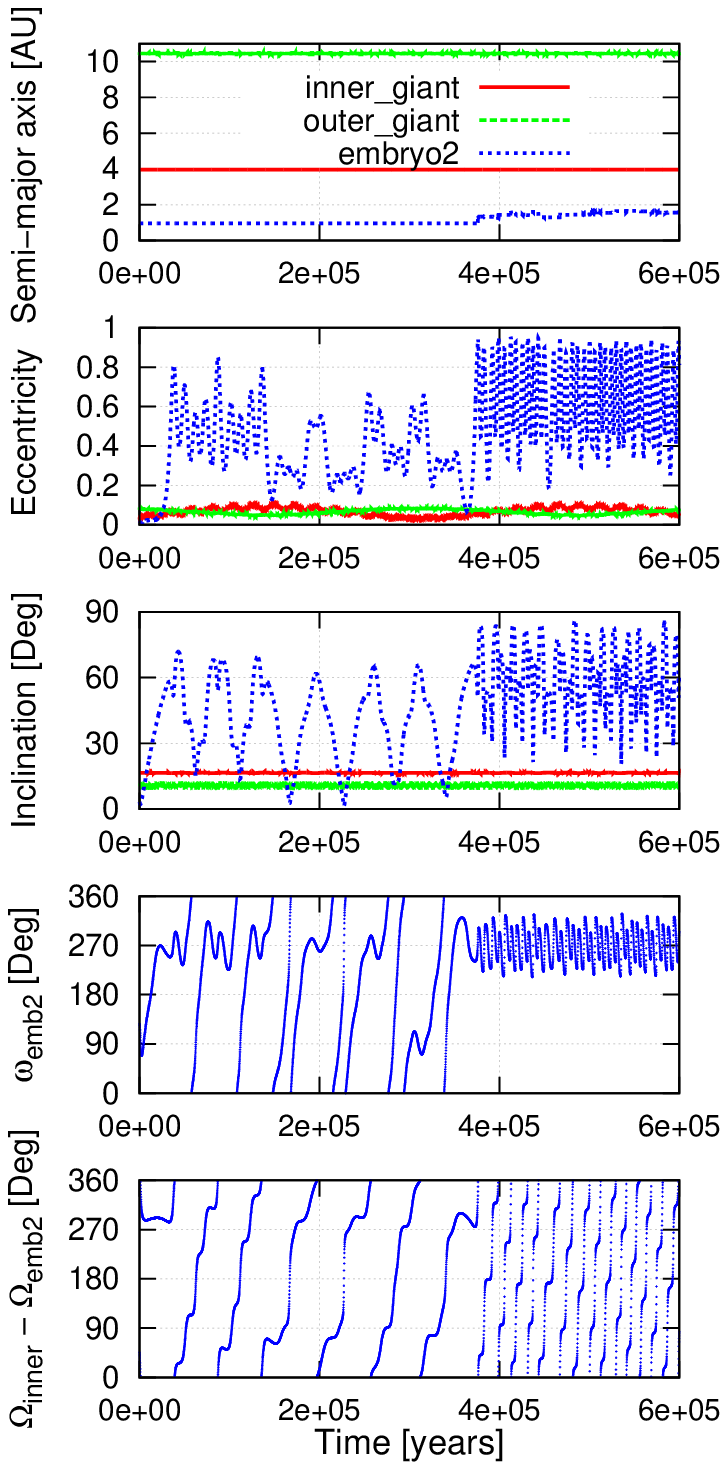}
        \includegraphics[width=0.48\hsize]{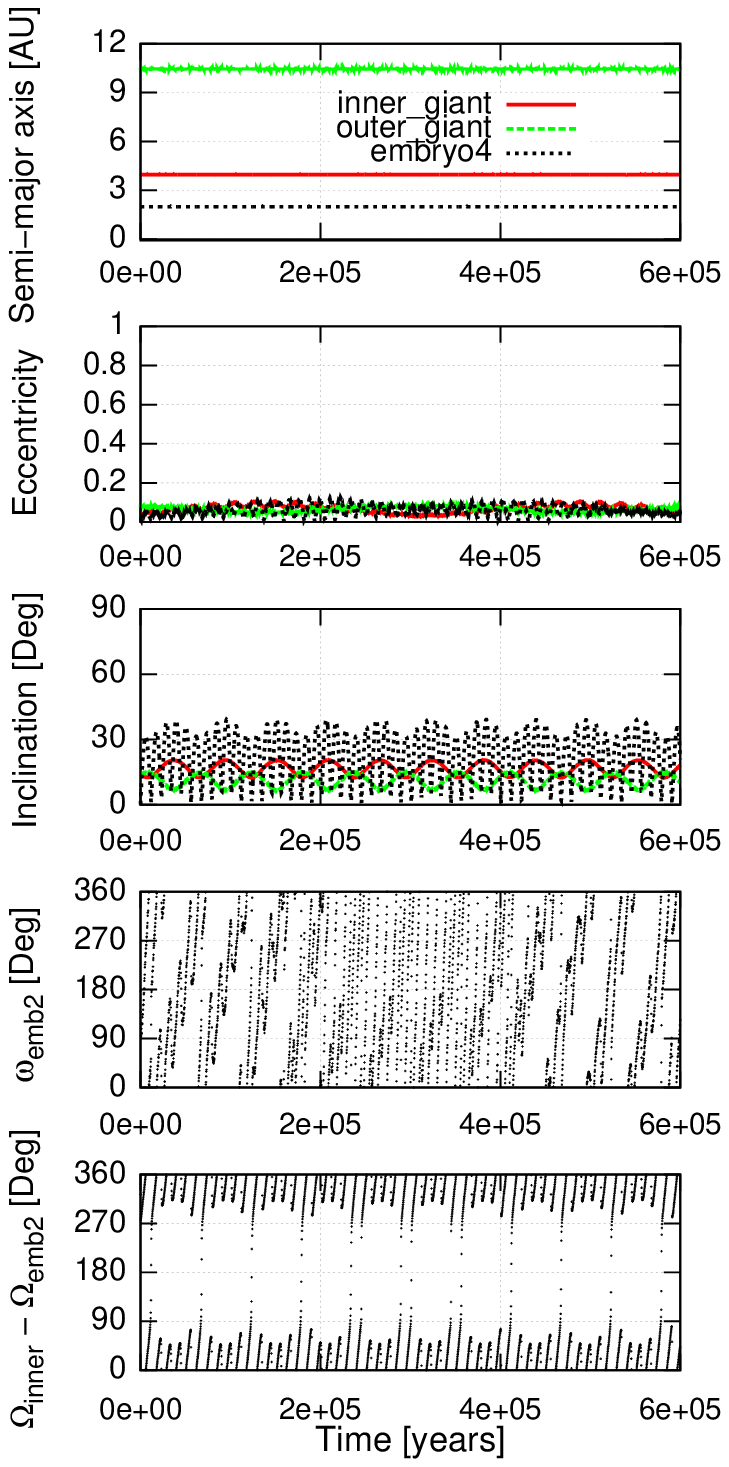}
       \caption{Long-term evolution of \texttt{embryo2} and \texttt{embryo4} of Fig.~\ref{fig:Kozai2}. The last two panels show the resonant angles associated to the Kozai and nodal resonances. See text for more details.}
       \label{fig:Kozai3}
 \end{figure}

To investigate the origin of the inclination waves, we report in Fig.~\ref{fig:Kozai2} an experiment of a simplified version of the \texttt{2\_05} system, consisting of only four embryos, initially located at $0.7$, $1$, $1.5$ and $2$ AU. Two different evolutions are visible. For \texttt{embryo1} and \texttt{embryo2} ($a_1=0.7$ and $1$, respectively), the first increase of the inclinations is due to a nodal resonance, as previously mentioned by \citet{LevisonAgnor}. As shown in Fig.~\ref{fig:Kozai3} (left panel), the difference between the longitude of the node of the inner giant and the one of \texttt{embryo2}, $\Delta \Omega$, oscillates around $270^{\circ}$ during the first $30000$ yr  (bottom panel). It leads to an increase of the inclination up to a value large enough for the embryo to be influenced by the Lidov-Kozai mechanism, but no eccentricity excitation. When the embryo inclination is close to $~40^\circ$ (the inclination value depends on the ratio of the semi-major axes between the embryo and the concerned giant planet), the systems can be captured in the Lidov-Kozai mechanism, in which the argument of the perihelion of the embryo, $\omega$, librates and its eccentricity and inclination undergo large amplitude variations. The evolution of \texttt{embryo2} shows several alternative phases of oscillation of $\omega$ around $90^\circ$ or $270^\circ$ and oscillation of $\Delta \Omega$, explaining the irregular evolution of the eccentricity observed in Fig.~\ref{fig:Kozai3} (left panel). At $0.4$ Myr, \texttt{embryo2} is finally captured in the Lidov-Kozai mechanism.

 \begin{figure}
       \centering
       \includegraphics[width=\hsize]{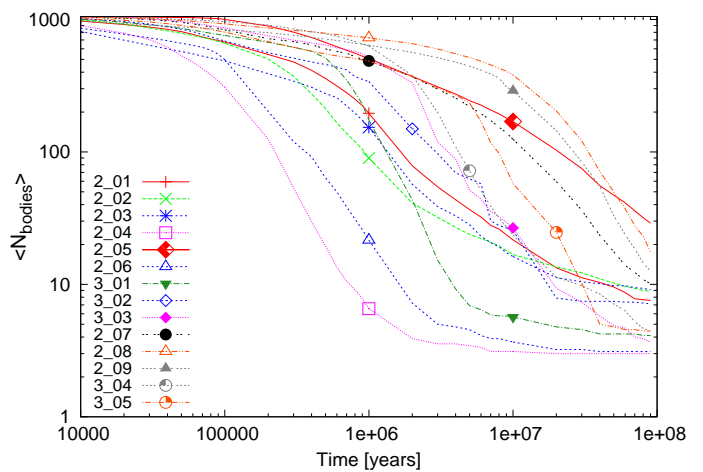} 
       \caption{Average number of bodies versus time, for all the nine runs of the fourteen giant planet configurations.}
       \label{fig:fig2}
 \end{figure}

 \begin{figure*}
       \centering
       \includegraphics[width=\hsize]{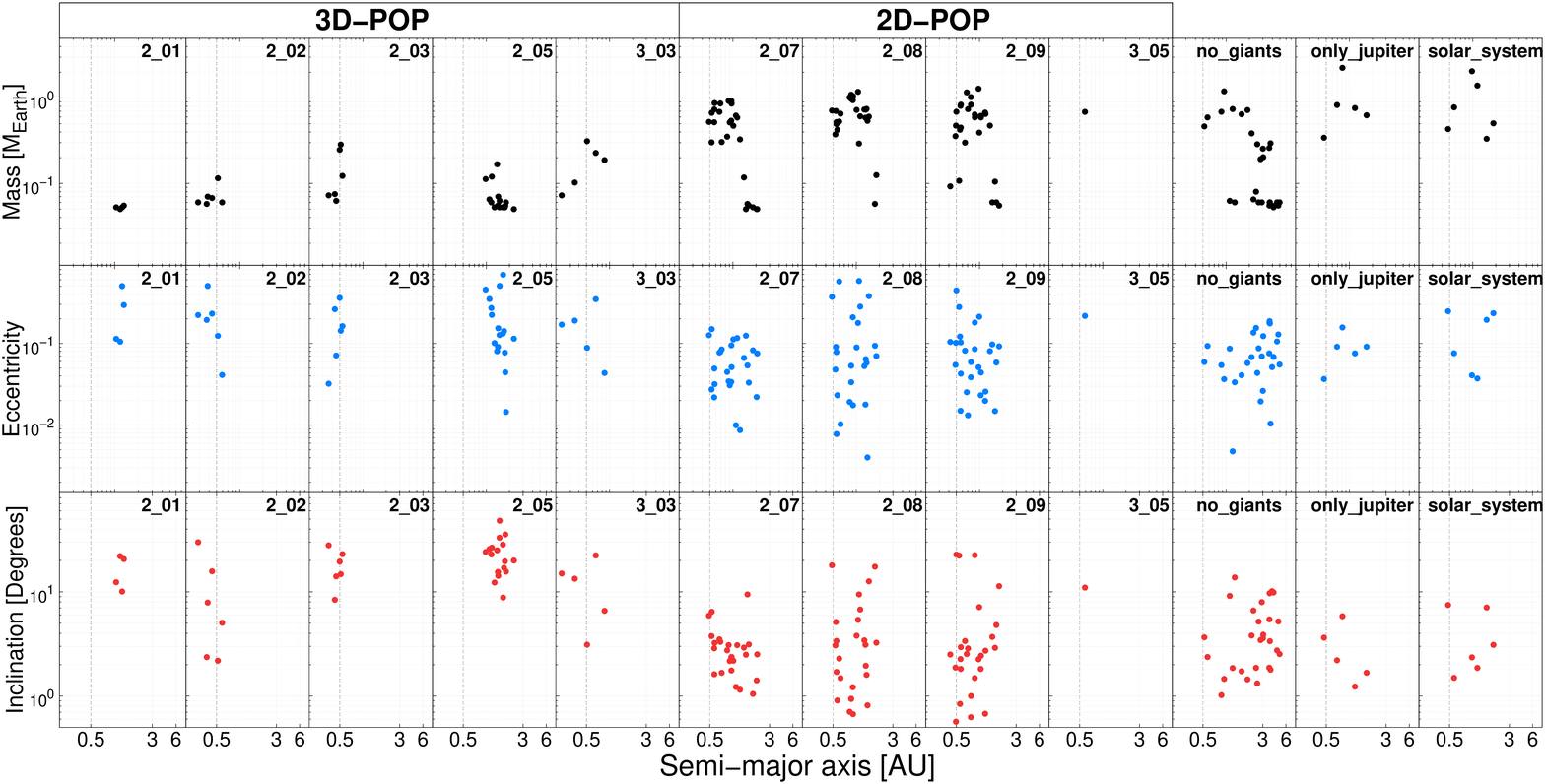} 
       \caption{For each giant planet configuration, mass (top panels), eccentricity (middle panels) and inclination (bottom panels) of the $116$ planets formed in our simulations (at $100$ Myr), as a function of the semi-major axis. In the three right-hand panels, the additional reference sets of simulations are shown for comparison: \texttt{no\_giants}, \texttt{only\_Jupiter} and \texttt{Solar\_System} (see the text for more details).}
       \label{fig:all}
 \end{figure*}

A second behavior is observed for \texttt{embryo3} and \texttt{embryo4}, with no increase of the eccentricities. As shown by Fig.~\ref{fig:Kozai3} (right panel), the inclination of \texttt{embryo4} has a periodic variation with moderate amplitude driven by a nodal resonance, as previously. However, the value reached during the secular variations of inclination are not large enough for the Lidov-Kozai mechanism to settle down.         

We conclude that the inclination waves observed in Fig.~\ref{fig:Kozai1} is a consequence of the large inclination variations of the embryos caused by the nodal resonance or the Lidov-Kozai mechanism, each particle having a different semi-major axis and thus a different amplitude in the inclination variation.


\section{Characterizing the formed terrestrial planets}\label{section5}
	First we give an overall overview of our simulations and the diversity of the outcomes, before describing in detail the parameters of the terrestrial planets formed in each system configuration.

\begin{table}[t!] 
\begin{center}

  \caption{Nature of the discard events of planetesimals and embryos. For the nine runs per configuration (index in the first column), the average percentage of collisions with a giant or the star is given in the second column and the one of ejections in the third column. The fourth column shows the average final eccentricity of the innermost giant planet for the nine runs per configuration, while the minimum and maximum values are given in parenthesis. The last column indicates the average remaining mass (in $M_{\Earth}$) of the terrestrial disc after $100$ Myr, while in parenthesis is indicated the maximum mass of the formed terrestrial planets in each set.} \label{tab:table1} 

   \tiny

 \begin{tabular}{ l | l | l | l | l}
 
\hline
	      &    Accreted      & 			 & Final inner		& Remaining  mass	    \\  
   Index  &    by giants     & Ejected    & eccentricity 		& in $M_{\Earth}$ 	    \\ 
          &    or the star   &     	    & {(min - max)}  	   	& (max)			\\ \hline
 \texttt{2\_01}  &    51.9\% &  48.1\%   & 0.228 \,{(0.132-0.291)}	&   0.033  \,\,{(0.055)} \\ 
 \texttt{2\_02}  &    44.3\% &  55.7\%   & 0.143 \,{(0.034-0.196)}	&   0.061  \,\,{(0.115)} \\ 
 \texttt{2\_03}  &    43.6\% &  56.4\%   & 0.385 \,{(0.186-0.528)}	&   0.110  \,\,{(0.285)} \\ 
 \texttt{2\_04}  &    82.6\% &  17.4\%   & 0.430 \,{(0.311-0.553)}	&   0.000  
 \\     
 \texttt{2\_05}  &    57.9\% &  42.1\%   & 0.070 \,{(0.026-0.194)}	&   0.183  \,\,{(0.168)}	 \\ 
 \texttt{2\_06}  &    65.5\% &  34.5\%   & 0.406 \,{(0.203-0.716)}	&   0.0003 
 \\ 
 \texttt{3\_01}  &    47.0\% &  53.0\%   & 0.410 \,{(0.087-0.855)}	&   0.000  
 \\ 
 \texttt{3\_02}  &    43.2\% &  56.8\%   & 0.434 \,{(0.186-0.784)}	&   0.009  \,\,{(0.580)}	 \\
 \texttt{3\_03}  &    90.9\% &   9.1\%   & 0.329 \,{(0.081-0.823)}	&   0.101  \,\,{(0.313)}	\\\hline    
 \texttt{2\_07}  &    37.0\% &  63.0\%   & 0.155 \,{(0.077-0.279)}	&   1.348 \,\,{(0.925)}	 \\     
 \texttt{2\_08}  &	 8.8\%  &  91.2\%   & 0.105 \,{(0.041-0.186)}	&   1.851 \,\,{(1.180)}	 \\
 \texttt{2\_09}  &	 22.2\% &  77.8\%   & 0.160 \,{(0.064-0.384)}	&   1.626 \,\,{(1.283)} 	 \\
 \texttt{3\_04}  &	 51.8\% &  48.2\%   & 0.331 \,{(0.139-0.774)}	&   	0.023 \,\,{(0.198)}   	 \\
 \texttt{3\_05}  &	 33.1\% &  66.9\%   & 0.233 \,{(0.097-0.569)}	&   0.099 \,\,{(0.688)}	 \\
  \hline  
 \end{tabular}
   
\end{center}
\end{table}

 \begin{figure}
       \centering
       \includegraphics[width=0.75\hsize]{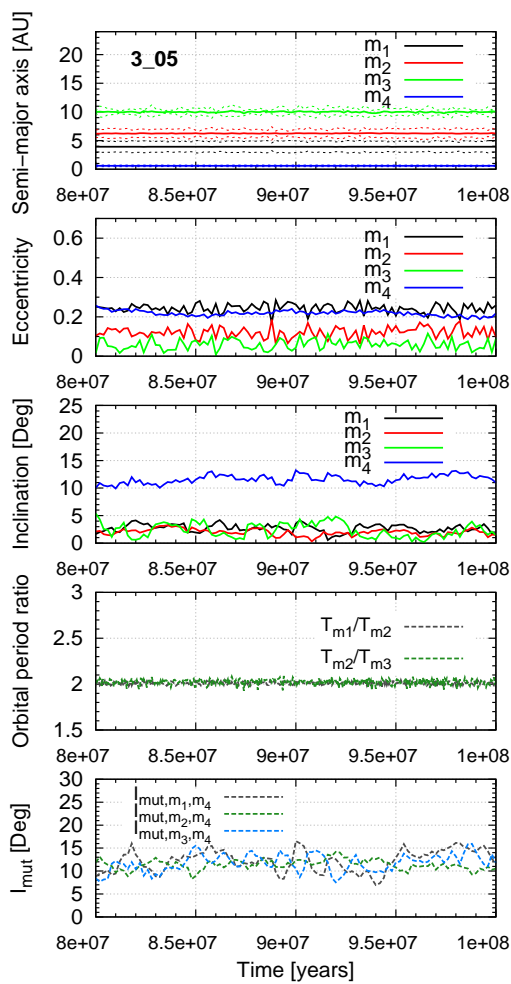} 

       \caption{Formation of an inclined terrestrial planet in coplanar system (2D-POP, \texttt{3\_05}), consisting of three giants in Laplace resonance. The notations $m_{1}, m_{2}$ and $m_{3}$ refer to the inner, middle and outer giants, respectively, and $m_{4}$ to the terrestrial planet. The 1:2:4 resonance is preserved throughout the evolution of the system. In the top panel, the periastron and apoastron of the planets are displayed with dashed lines.}
       \label{fig:case4}
 \end{figure}

	\subsection{Remaining terrestrial mass}\label{remaining}
	
	The average number of bodies over time is shown in Fig.~\ref{fig:fig2}, for the nine runs of each configuration. We observe that, for most of the systems, it only slightly varies after $50$ Myr. In some configurations, such as \texttt{2\_04} and \texttt{2\_06}, nearly all the terrestrial bodies from the disc are even discarded within a few million years, due to the strong perturbations exerted by the giants. Let us note that much later instabilities, on timescales longer than $100$ Myr, are common in systems with eccentric Jupiter-like planets, as shown by \cite{2017Icar..288...88C}. 
	
	In Table~\ref{tab:table1}, we give more details on the nature of the discard events observed in our simulations and report, for each configuration, the average percentages of collisions with a giant or the star (second column) and ejections of the system (third column). It is clear that co-planar two-planet systems, like \texttt{2\_08} (Fig.~\ref{fig:case2}) and \texttt{2\_09} (Fig.~\ref{fig:case3}), give rise to massive ejections of planetesimals and embryos, due to the secular or resonant interactions with the giant planets. Lots of collisions with the star are reported for systems in 3D-POP as a result of the Kozai secular excitation of the disc by inclined giant planets, as described in Section~\ref{section4}.
	
	The last two columns of Table~\ref{tab:table1} show the average final eccentricity of the innermost giant and the average remaining mass of the terrestrial bodies, at the end of the simulations for the nine runs of each configuration. Our results indicate that there is a correlation between the final eccentricity of the inner giant and the total terrestrial mass at $100$ Myr, as previously noted by \citet{Raymond2011}. The more eccentric the innermost giant planet, the less efficient the terrestrial accretion process. Furthermore, only the two-planet configurations of 2D-POP (\texttt{2\_07}, \texttt{2\_08} and \texttt{2\_09}) have an average remaining mass above $1 M_{\Earth}$, showing that the accretion of terrestrial planets is more efficient in coplanar two-planet systems than in non-coplanar systems or systems with three giant planets. This difference of evolution also has an impact on the parameters of the formed terrestrial planets, as we will show in the following.

	\subsection{Diversity of the terrestrial planets}
	
	In the $126$ simulations, we have formed a total of $116$ terrestrial planets with mass $>0.05 M_{\Earth}$, gathered in $54$ systems. All the planets are reported in Fig.~\ref{fig:all}, which shows, for each configuration of giant planet system, the masses (top panels), the eccentricities (middle panel), and the inclinations (bottom panels) of the terrestrial planets formed, as a function of their semi-major axis. It is interesting to note that the planets formed in the different runs of each configuration are rather similar, but they differ quite substantially between 2D-POP and 3D-POP. Indeed the planets formed in 2D-POP are more numerous, and their parameters more various than the ones of 3D-POP. The last three columns of Fig.~\ref{fig:all} show the terrestrial planets obtained in the three reference cases discussed in Section~\ref{primary}. The planets formed in the \texttt{no\_giant} set-up look very similar to the ones of 2D-POP, which suggests that the giant planets of 2D-POP have a rather limited impact on the accretion process. The diversity of terrestrial planets is mainly due to the interactions between the planetesimals, embryos and terrestrial planets themselves. On the contrary, the two other additional sets of simulations (\texttt{only\_Jupiter} and \texttt{Solar\_System}) impose strong constraints on the parameters of the terrestrial planets. The planets formed in these two sets are all rather identical, as it is the case in the simulations of 3D-POP.
	
	Moreover, we also observe that two-planet configurations are more efficient in planet formation than three-planet systems. Terrestrial planets in three-planet systems are only found in the non-coplanar \texttt{3\_03} architecture (5 terrestrial planets in total) and the coplanar \texttt{3\_05} architecture (1 in total). Nearly all these planets are inclined, even in the 2D-POP as shown in Fig.~\ref{fig:case4}, where the evolution of the unique planet formed in the simulations of the \texttt{3\_05} system is displayed. While the three giant planets approximately share the same orbital plane, the terrestrial planet evolves on an orbital plane whose inclination is about $12^\circ$.

 \begin{figure}
       \centering
       \includegraphics[width=\hsize]{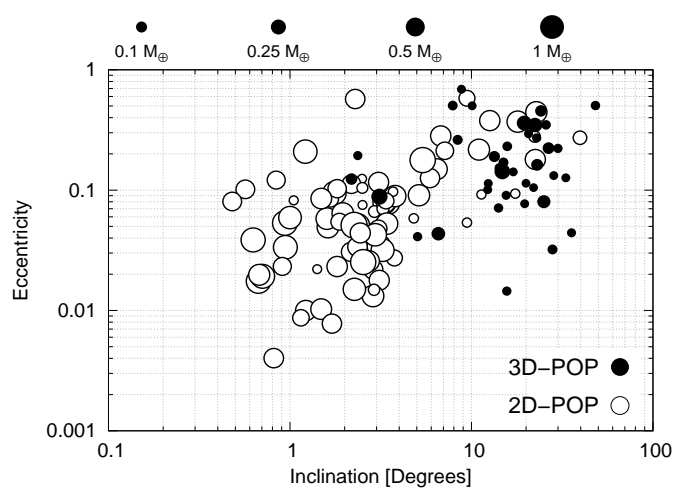} 
       \caption{Eccentricity versus inclination of the terrestrial planets. White circles correspond to planets formed in coplanar giant planet systems (2D-POP) and black circles to planets formed in 3D systems (3D-POP). The size of each circle is proportional to the cubic root of their mass. We observe that all the bodies that have formed  after 100 Myr, evolve in orbits with $e<0.3$ and $i<10^{\circ}$. }
       \label{fig:figINC_VS_ECC}
 \end{figure}

 \begin{figure}[t!]
       \centering
       \includegraphics[width=\hsize]{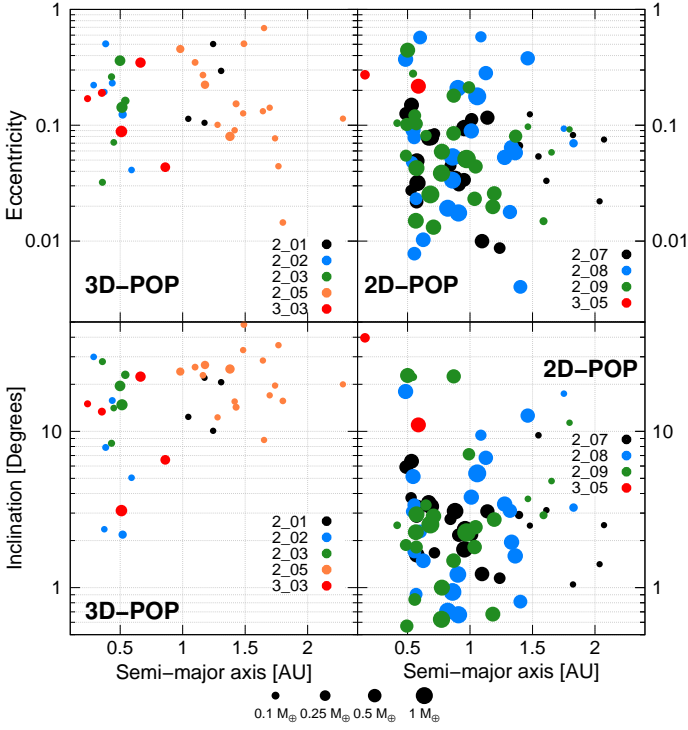} 
       \caption{ \textit{Top row}: Eccentricity versus semi-major axis for the planets formed in 3D-POP (left) and 2D-POP (right). \textit{Bottom row}: Inclination versus semi-major axis for 3D-POP (left) and 2D-POP (right). Vertical axes are on logarithmic scale. The size of each circle is proportional to the cubic root of the planetary mass.}
       \label{fig:figA_VS_ECC_AND_INC}
 \end{figure}

	\subsection{Physical and orbital parameters}

	The eccentricities, inclinations and masses of the terrestrial planets formed in our simulations are displayed in Fig. \ref{fig:figINC_VS_ECC}. As previously, the size of each circle is proportional to the cubic root of the planetary mass. The white and black circles represent the planets from 2D-POP and 3D-POP, respectively. As expected, there is a stark contrast between the two populations. While the majority of the terrestrial planets in 2D-POP are massive and on low-eccentric and low-inclined orbits, similarly to what has been observed for the three reference cases in Fig. \ref{new2}, the planets of 3D-POP are generally less massive, with larger eccentricities and inclinations. This is a direct consequence of the different excitation mechanisms acting on the disc of planetesimals and embryos highlighted in Section \ref{remaining}. Due to secular and resonant perturbations acting both in the inner and outer discs, 3D-POP systems suffer from more discard events, and the accretion is thus less efficient for non-coplanar systems.    

\begin{figure}
       \centering
       \includegraphics[width=0.49\hsize]{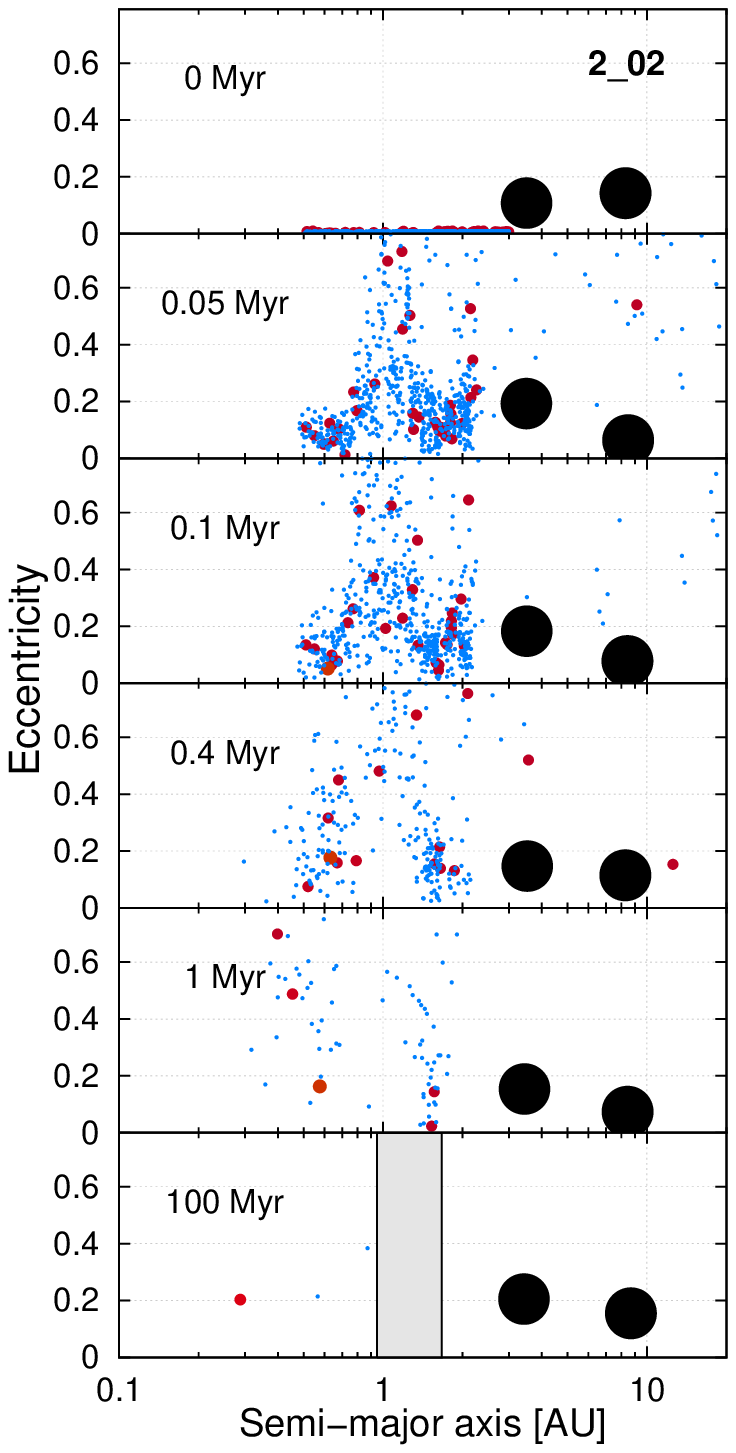}
       \includegraphics[width=0.49\hsize]{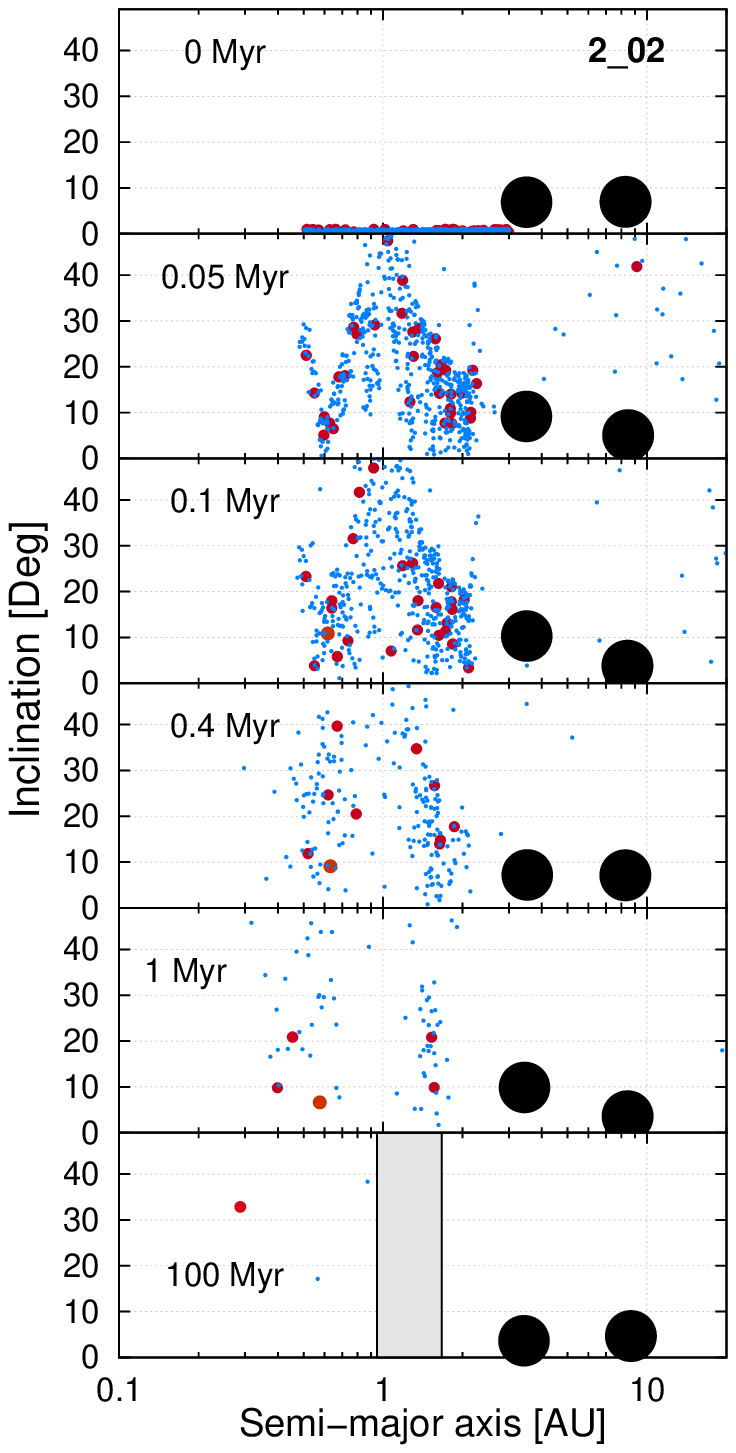}
       \includegraphics[width=0.95\hsize,keepaspectratio] {figures/colorbar.eps}
       \caption{Snapshots in time from a non-coplanar system (3D-POP, \texttt{2\_02}) whose evolution results in the formation of a terrestrial planet closer to the star than the inner edge of the disc. Formatted as in Fig.~\ref{fig:case1}.}
       \label{fig:ex_inner}
 \end{figure}

	Information on the proximity to the star of the terrestrial planets formed in 2D-POP and 3D-POP is given in Fig. \ref{fig:figA_VS_ECC_AND_INC}, which displays the eccentricity (top panels) and inclination (bottom panels) of the planet as function of the semi-major axis. Again the planet mass is represented by the size of the circle. The planets formed in 2D-POP systems are nearly all located beyond the inner edge of the disc ($0.5$ AU). However, in 3D configurations, several planets are found closer to the star, at around $0.2-0.3$ UA. They result from scattering in systems where the secular and resonant perturbations, especially the Kozai excitation, are affecting nearly the entire disc. An example of such an evolution is displayed in Fig.~\ref{fig:ex_inner}.  

\begin{figure}
       \centering
       \includegraphics[width=\hsize]{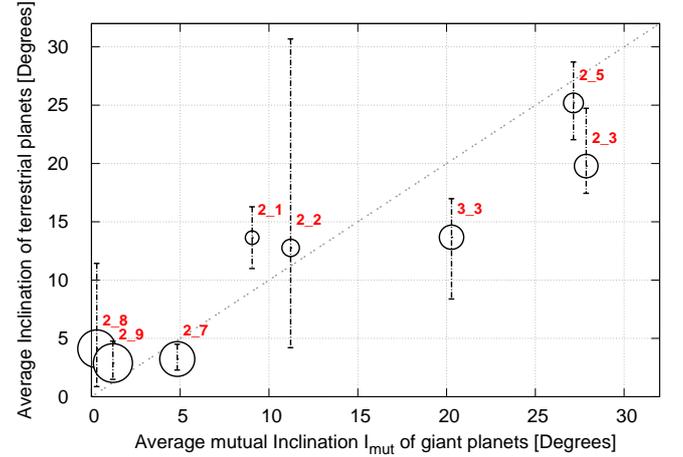} 
       \caption{Average inclination (with respect to the initial midplane of the disc) of all the terrestrial planets formed in two-planet systems. The horizontal axis shows the average mutual inclination between the two giants. The averaging values are computed over the last $10$ Myr of integration. For each configuration, the vertical error bars show the maximum and minimum inclinations of the terrestrial planets. The size of each circle corresponds to the largest planet found in a configuration and is proportional to the cubic root of the planetary mass.}
       \label{fig:meanINC}
 \end{figure}

	We examine in Fig.~\ref{fig:meanINC} the relation between the mutual inclination of the giant planets and the orbital inclination of terrestrial planets formed in two-planet configurations. The vertical axis indicates the average inclination with respect to the initial disc plane, of the terrestrial planets of each giant planet configuration (identified by its index next to each circle) over the last $10$ Myr. The vertical dashed lines correspond to the minimum and maximum inclinations observed in a configuration. On the horizontal axis, we indicate the average mutual inclination of the two giant planets. The size of each circle represents the largest terrestrial planet found in each architecture and is proportional to the cubic root of its mass. The more mutually inclined the giants are, the more inclined the terrestrial bodies. The linear trend is obvious in Fig.~\ref{fig:meanINC}, showing that the inclination of a terrestrial planet reflects the mutual inclination of its companion giant planets, at least for our case studies. 
	
\begin{figure}[t!]
       \centering
       \includegraphics[width=\hsize]{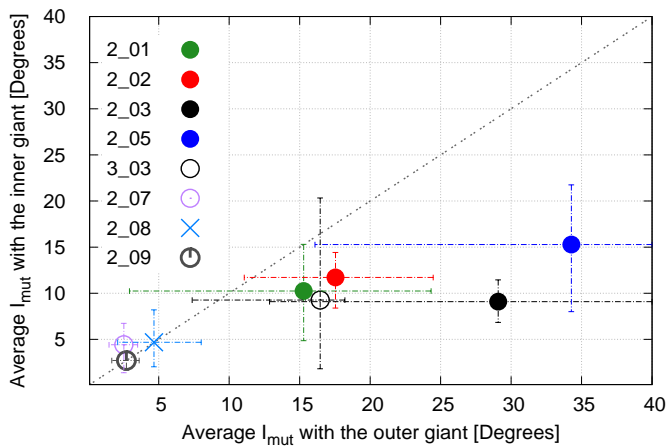} 
       \caption{For two-planet systems, average mutual inclinations between the terrestrial planets and the two giant planets. Mutual inclination with the inner giant planet is shown by the vertical axis, and mutual inclination with the outer planet by the horizontal axis. The averaging values are computed over the last $10$ Myr of integration. Error bars indicate the minimum and the maximum values of each giant-planet configuration.}
       \label{fig:inner_vs_outer3}
 \end{figure}

	Finally, for the two-planet configurations, we also show in Fig.~\ref{fig:inner_vs_outer3} the average mutual inclinations between the formed terrestrial planets and both giants over the last $10$ Myr of integration. Horizontal and vertical axes correspond to the mutual inclinations with the outer giant planet and the inner giant planet, respectively. The error bars indicate the minimum and maximum values in each configuration. We observe that terrestrial planets clearly evolve closer to the plane of the inner giant planet, whatever the mass ratio of the giants.
	
\section{Discussion and conclusions}\label{section6}

	In this study, we examined the formation of terrestrial planets in $14$ different giant planet systems. We began our simulations from the late-stage accretion phase, where planetary embryos and planetesimals interact each other gravitationally under the influence of the gas giants. The physical and orbital parameters of the giant planet systems considered in the present work result from n-body simulations of three giant planets in the late stage of the gas disc, under the combined action of Type II migration and planet-planet scattering, following \citet{Sotiriadis2017}. 
	
	We selected $9$ representative two- and three-planet systems where the giants, usually on eccentric orbits, have a mutual inclination larger than $10^\circ$ (3D-POP) and $5$ two- and three-planet coplanar configurations (2D-POP). We performed $9$ runs for each giant planet architecture. Our goal was to analyze the impact of these eccentric and inclined massive giant planets on the terrestrial planet formation process and investigate whether it can possibly lead to the existence of inclined terrestrial planets around solar-mass stars.

	Our simulations suffers from some limitations. First, we assumed that there is no gas/dust disc left in the systems and the giant planets are fully formed. Of course, the late-protoplanetary disc phase and the late-accretion phase are not independent of each other. The interactions between the two phases were not taken into account in this work for computational reasons. Due to this limitation, we have not included in our modelling the phenomena related to planet-disc interactions, such as orbital migration and eccentricity/inclination damping, for both gaseous and rocky bodies. 
	
	Secondly, the disc of planetesimals and planetary embryos, with the near-circular and near-coplanar orbits considered in our simulations, did not contain the imprint of the ``late-gas'' phase from where we acquired our initial set-up. The evolution history of the newborn giant planets affects the orbits of the terrestrial bodies that are present in the system. In particular, in such "exotic" giant planet systems, as the one we embraced for our initial conditions, planetesimals and embryos should have been excited in eccentric/inclined orbits much before the dispersal of the gas, especially if they have grown to be quite massive before the formation of the gas giants. Nevertheless, it would be computationally hyper-expensive and difficult to perform a large and reliable statistical ensemble of simulations including both phases, the "late-gas" and the "late-accretion" phase, and conclude about the formation of terrestrial planets in such non-coplanar frame. For this reason, we leave for future work a more realistic study that includes the joint evolution of giant planets and terrestrial bodies during the late stage of the protoplanetary disc and their interactions with the disc (Type-II migration, gas drag for planetesimals, etc). 
	
	Thirdly, the efficiency of planet accretion depends on several free parameters of our model. The initial disc mass, the total number of embryos and planetesimals in the disc, their total mass ratio and the inner and outer edge of the disc are some of the parameters that we keep constant in our ensemble of simulations. Also, the initial conditions of embryos and planetesimals are fixed in all of our simulations. \citet{2006ApJ...642.1131K} showed that the final structure of planetary systems depends only to a slight extent on the initial conditions and the distribution of protoplanets as long as the total mass is a fixed parameter. In addition, our assumption of an equal mass in planetesimals and embryos is based on \citet{2006AJ....131.1837K}, but the exact ratio might affect the timing of growth to some degree \citep{Jacobson20130174}. Moreover, \citet{2010ApJ...714L..21K} discussed the limitations of the perfect accretion model. They argued that despite only half of collisions could lead to accretion in a realistic accretion model, the final number of planets and their orbital architecture are barely affected by the accretion condition. Finally, fragmentation is not taken into account in our work \citep{2014Icar..233...83C}.

	In our simulations, we observed that accretion of terrestrial planets is more efficient in coplanar two-planet systems than in non-coplanar systems or systems with three giant planets. In these 2D architectures, starting initially of a 5 $M_{\Earth}$ disc of rocky bodies, the average remaining mass (on the 9 runs per system) is above $1 M_{\Earth}$ and the formation of a massive terrestrial body (m > 0.5  $M_{\Earth}$), inside the habitable zone, is very likely to happen (see Fig.~\ref{fig:case2}). Moreover, one or several terrestrial planets are formed in 2D-POP systems, and they are usually evolving in low-eccentricity and low-inclination orbits. Nevertheless, Earth-like planets could also emerge in stable inclined orbits even when the gas giants are evolving in a coplanar configuration (see Fig.~\ref{fig:case4}). Concerning the systems of 3D-POP, fewer terrestrial planets are formed compared to the coplanar architectures. The influence of mutually inclined and eccentric giant planets is strong and the dynamical excitation of the planetesimals and embryos occurs on a very short timescale, driven by resonant and secular interactions with the giants. In particular, the Lidov-Kozai mechanism strongly affects the disc of planetesimals and embryos, by inducing eccentricity and inclination waves in the first thousand years of the simulations. Most of the rocky material is either accreted by the central star of ejected from the system during the first few million years and as a consequence less massive bodies are formed at the end of integration. Another important outcome is that the terrestrial planets formed in 3D-POP are found on eccentric and inclined orbits, the inclination of the terrestrial planets being generally similar to the mutual inclination of the giant planets. As a result, we stress that the formation of terrestrial planets on stable inclined orbits is possible through the classical accretion theory, both in coplanar and non-coplanar giant planet systems.

\begin{acknowledgements}
This work was supported by the Fonds de la Recherche Scientifique-FNRS under Grant No. T.0029.13 (“ExtraOrDynHa” research project) and by the Hubert Curien/Tournesol Program. Computational resources have been provided by the Consortium des Équipements de Calcul Intensif (CÉCI), funded by the Fonds de la Recherche Scientifique de Belgique (F.R.S.-FNRS) under Grant No. 2.5020.11. S. S. thanks the LAB for their hospitality during his stay in Bordeaux. S. N. R. thanks the Agence Nationale pour la Recherche for support via grant ANR-13-BS05-0003-002 (project MOJO).
\end{acknowledgements}

\bibliographystyle{aa}
\bibliography{bibliography}

\end{document}